\newcommand{\Pl}{\partial}
\newcommand{\ts}{\textstyle}
\newcommand{\bee}{\begin{equation}}
\newcommand{\ene}{\end{equation}}
\newcommand{\beea}{\begin{eqnarray}}
\newcommand{\enea}{\end{eqnarray}}

\newcommand{\fpar}[2]{\frac{{\ts \Pl \/ #1}}{{\ts \Pl \/ #2}}}

\newcommand{\npar}[3]{\frac{{\ts \Pl^{#1} \/ #2}}{{\ts \Pl \/ #3^{#1}}}}

\documentclass[aps,prb]{revtex4-1}
 \usepackage{graphicx}
 \usepackage{dcolumn}
 \usepackage{bm}
 \usepackage{latexsym}
 \usepackage{amsmath}
 \usepackage{amssymb}
 \usepackage{amsfonts}
 \usepackage{amsthm}
 \begin{document}
 \title{Nonlinear  Shear Wave in a Non Newtonian Visco-elastic Medium}
 \author{D. Banerjee M. S. Janaki and N.  Chakrabarti }
 \affiliation{ Saha Institute of Nuclear Physics,
 1/AF Bidhannagar Calcutta - 700 064, India.}
 \author{M. Chaudhuri}
\affiliation{ Max-Planck-Institut f\"{u}r extraterrestrische Physik, 85741 Garching, Germany}
\begin{abstract}
      An analysis of nonlinear transverse shear wave has been carried out on non-Newtonian viscoelastic liquid using generalized hydrodynamic(GH) model. 
The nonlinear viscoelastic behavior is introduced through velocity shear dependence of viscosity coefficient by well known Carreau -Bird model. 
The dynamical feature of this shear wave leads to the celebrated Fermi-Pasta-Ulam (FPU) problem. Numerical solution has been obtained which shows 
that initial periodic solutions reoccur after passing through several patterns of periodic waves. A possible explanation for this periodic solution 
is given by constructing modified Korteweg de Vries (mKdV) equation. This model has application from laboratory to astrophysical plasmas as well as 
biological systems.
\end{abstract}
\maketitle

\section{Introduction}
\label{intro}
   Viscoelastic properties of fluids in general and dusty plasmas in particular, have acquired high attraction recently
   due to the experimental characterization of non Newtonian nature and understanding of the associated physical processes.
   The interplay between shear flow and viscosity in a non Newtonian fluid, are purely nonlinear,
   therefore conventional linear theories are unfit to predict the behavior that what we observe in a medium.
   In response to external stresses viscoelastic media  dissipate energy due to viscosity and store energy due to elastic nature.  An
   important manifestation of such elastic property is reflected in the collective
   behavior with the prediction of a large amplitude transverse shear wave requesting some nonlinear physics.
   \newline
   Typically charged particle systems can  be modeled as Yukawa systems that can have a liquid state \cite{shma,ichi,pkmt}
   these include colloids, dense astrophysical plasmas\cite{msjn}, strongly coupled dusty plasmas \cite{pkaw,mash} and magnetized plasma \cite{banj}.
   A Yukawa fluid possesses a memory dependent nonlocal viscoelastic coefficient enabling the propagation of
   a transverse shear wave that has been experimentally observed in a strongly coupled dusty plasma\cite{pram}. Another important characteristic
   exhibited by strongly coupled fluid is its non Newtonian behavior where the viscosity coefficient changes with velocity shear rate.
   This is also been observed in recent laboratory experiments \cite{nose,ivle,jpps}. 
    Comparable examples of viscoelastic, non Newtonian behavior are also found in other branches of physics
    such as complex fluid \cite{sora,mamu},
    polymeric liquids\cite{bird}, colloidal suspensions\cite{bisw}, human blood\cite{thur}.
 \newline
    In this work, we construct a strongly nonlinear shear wave equation, for a non Newtonian viscoelastic
    fluid and  find a possible solution of this nonlinear equation. The solution we found in a sense mimic the nonlinear solution of
    famous Fermi-Pasta-Ulam (FPU) problem\cite{fpu}. Numerically integrated discrete equation is expanded
    in Taylor series in continuous variable  and a dispersive effect is shown to exist and form a Modified Korteweg de Vries (mKdV) equation.
    The solution of mKdV equation helps us to explain the nonlinear periodic solutions.

         \section{Generalized Hydrodynamic Equations with Non-Newtonian Viscosity} \label{sec:bas}
         A strongly coupled dusty plasma whose constituents are electrons, ions and negatively fixed charged, massive dust
       grains are considered.  The dust grains are strongly correlated to each other  due to their larger
       electric charge and lower temperature, whereas, both electrons and ions are weakly coupled because of their smaller electric charges
       and higher temperatures. So the dynamics of shear waves in strongly coupled dusty plasma can be modeled  by a memory
       dependent viscoelastic operator involving relaxation time $\tau_m$ that leads to generalized hydrodynamic
       momentum equation \cite{fren,pkaw} given by
        \bee
        \left \{1+ \tau_m \left (\fpar{}{t} +  {\bf v}\cdot \nabla\right) \right \} \left[\rho \left(\fpar{}{t}+{\bf v}\cdot \nabla\right){\bf v} -
        {q n}{\bf E}+ \nabla p \right ] = \fpar{\sigma_{ij}}{x_{j}}
        \label{deq}
        \ene
        where  ${\bf v}$, $\rho$, $n$, $p$, ${\bf E}$ are respectively  fluid velocity, mass density, number density,  pressure,
         electric field and $\sigma_{ij}$  strain tensor of the  medium. In a standard approach  electrons and ions are light fluid and can be 
         connected through electric field. However, for electrostatic shear wave, electric field will have no role to play hence explicit
         electron and ion dynamics are not taken into account. For an incompressible medium strain tensor is given by
        \[
        \sigma_{ij} = \eta(S) \left(\frac{\partial v_{i}}{\partial x_{j}} + \frac{\partial v_{j}}{\partial x_{i}}\right).
        \]
        For a Newtonian fluid $\eta $ is generally constant. However, for a
        non-Newtonian incompressible fluid, it has been shown that \cite{bird},
        $\eta$ depends on the scalar invariants of strain tensor  which can be written as
         \[
         I = \sum_i \sum_j \left(\frac{\partial v_{i}}{\partial x_{j}}
         + \frac{\partial v_{j}}{\partial x_{i}}\right)\left(\frac{\partial v_{i}}{\partial x_{j}}
         + \frac{\partial v_{j}}{\partial x_{i}}\right).
         \]
         and $S = \sqrt{I/2}$.
         It has been shown that in the limit $\tau_{m}{\partial/\partial t} \gg 1 $,
         viscoelastic fluid can  support shear wave like transverse waves in elastic rods\cite{pkaw}.
         Since the motion is considered incompressible, $\rho$ remains constant in space and
         time and the velocity is the only dynamical variable. Our system of equations utilizes the geometry
         of shear waves, i.e. the $x$ and  $t$ dependent velocity field in $y$-direction,
         the wave propagation is in $x$- direction. In such situation
         strain tensor $\sigma_{ij}$ and parameter $S$ are given by
         $\sigma_{yx} = \eta(S) \partial v_y/\partial x$, $S=\partial v_y/\partial x$.
         Using these, equation (\ref{deq}) is simplified as
        \bee
        \npar{2}{v}{t} = \fpar{}{x} \left[ \eta(S) \fpar{v}{x}\right]
        \label{simp}
        \ene
        To write the above equation we have used  normalization as follows:
        $
        v \rightarrow {v_y}/{c_{sh}}$, $\eta \rightarrow {\eta}/{\eta_0}$, $x \rightarrow {x}/{L}$,
        $t \rightarrow {t c_{sh}}/{L}$
        where, $\eta_0, L$, are  some  arbitrary viscosity and length, $c_{sh} = \sqrt{\eta_0/\rho\tau_{m}}$.
        To proceed further one needs to know the functional form of viscosity. In order to specify $\eta$ we have taken some
        input from recent experimental results\cite{ivle}. In the experiment it was shown
        that charged fluids like dusty plasma have both shear thickening and thinning properties depending upon the parameter regime of shear rate.
        In order to include both properties in our solution we have taken  well known
        Carreau-Bird viscosity model \cite{bird}. There are other models but this  model has an added advantage
        in a sense that one can easily recover the Newtonian viscosity.  Mathematically this model is expressed as
         $
         \eta \left(S\right) = 1 + \alpha S^{2}.
         $
         The parameter $\alpha $ is used as a measure of non-Newtonian effect, and is assumed to be small in Carreau Bird model.
          Obviously for $\alpha $ negative(positive) the fluid behaves like a shear thinning(thickening) medium.
         \section{Nonlinear Shear Waves, Energy Sharing, Recurrence}\label{sec:ana}
        Introduction of this model into equation (\ref{simp}), we obtain
         \bee
         \npar{2}{v}{t} = \npar{2}{v}{x} + \alpha \fpar{}{x} \left( \fpar{v}{x}\right)^3
         \label{nonlin}
         \ene
         As it is mentioned before, in absence of non-Newtonian effect i.e. $\alpha=0$
         we get back the linear shear wave equation (normalized form)  which  resembles the  wave equation  in elastic media.
         The nonlinear wave equation (\ref{nonlin}) is characterized by a parameter $\alpha$ therefore the solution depends on a
         single independent external parameter.

\begin{figure}
           \centering
             \begin{tabular}{cc}
                \resizebox{40mm}{!}{\includegraphics{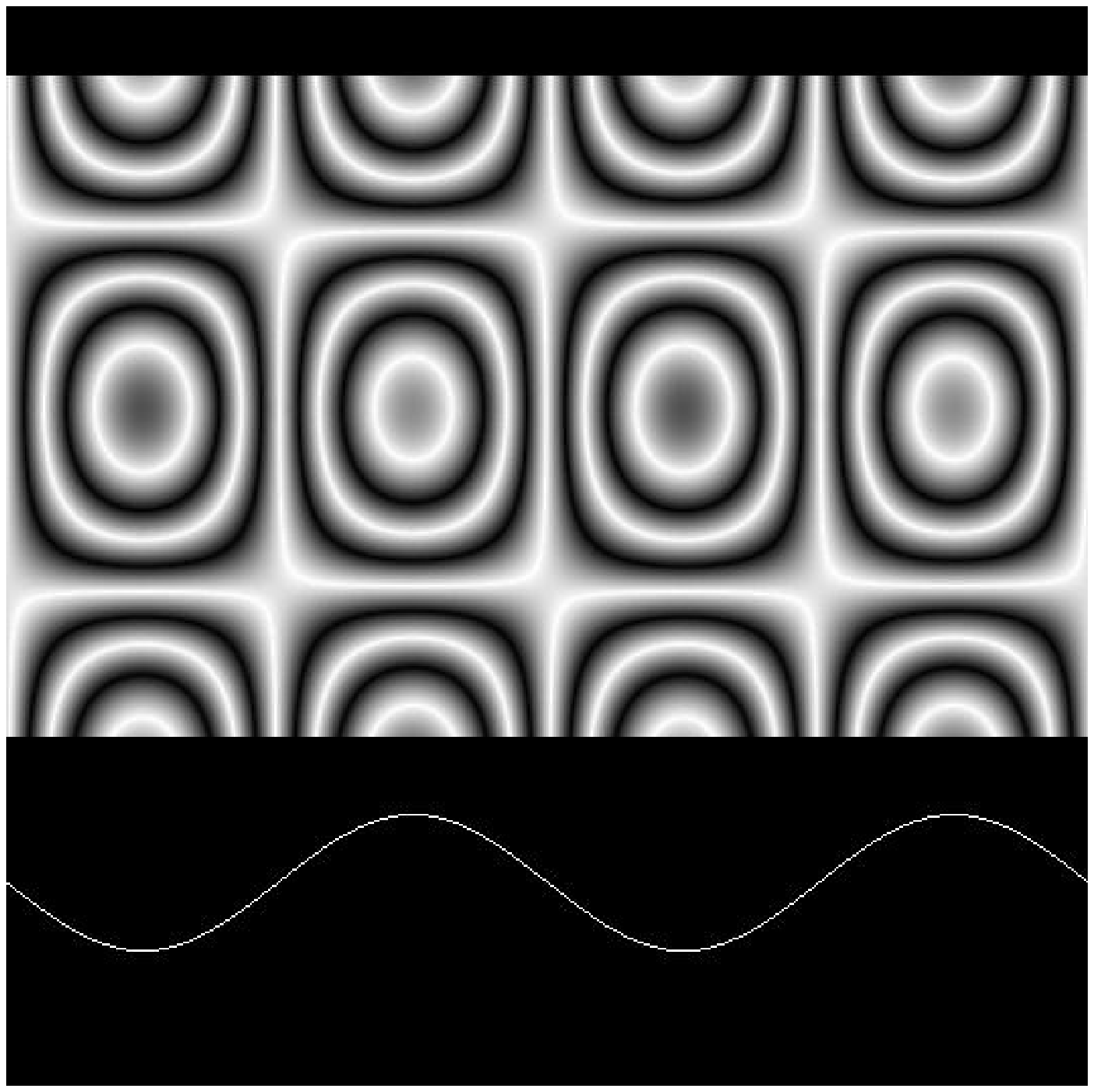}}&\resizebox{40mm}{!}{\includegraphics{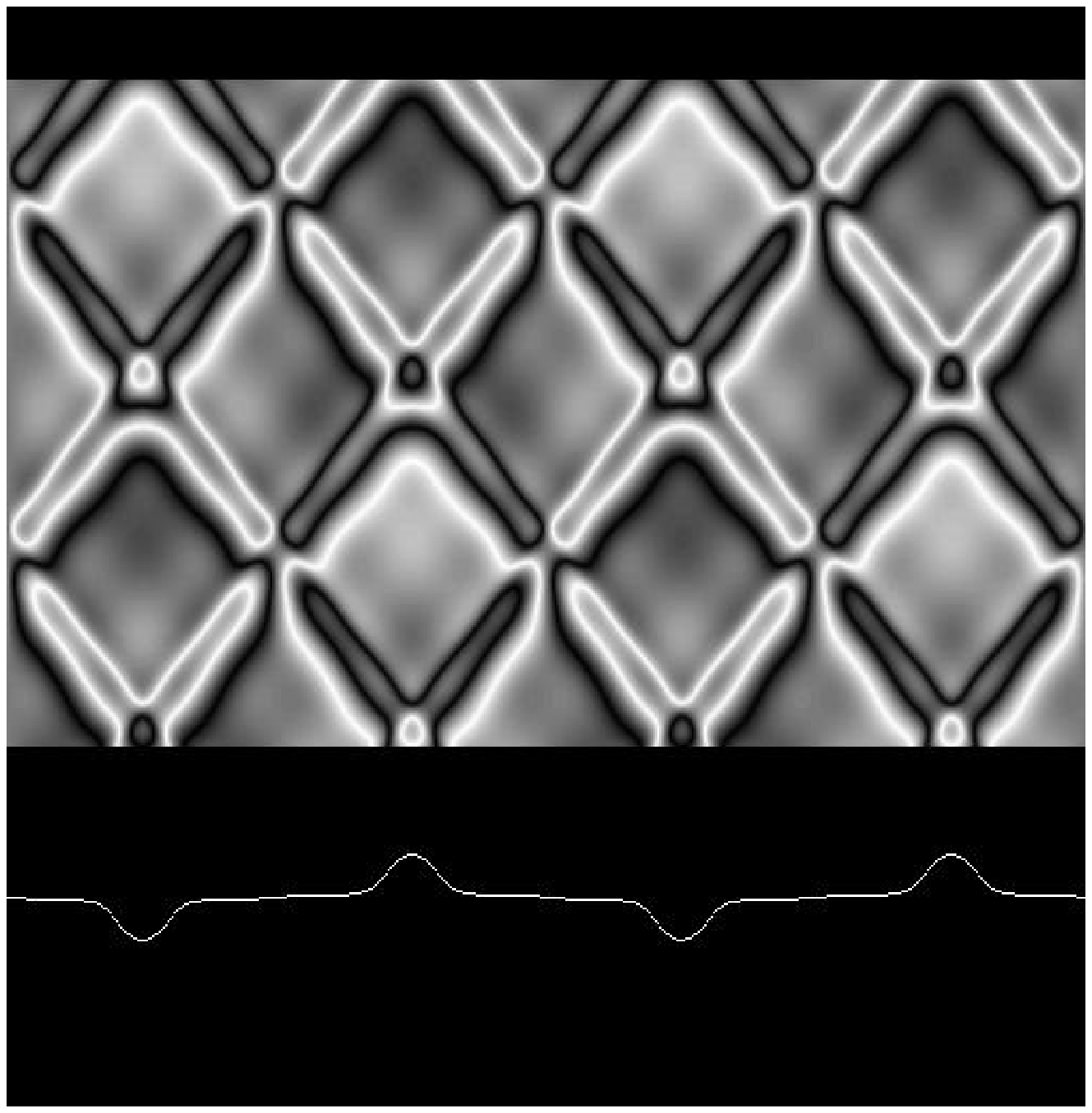}}\\
                \resizebox{40mm}{!}{\includegraphics{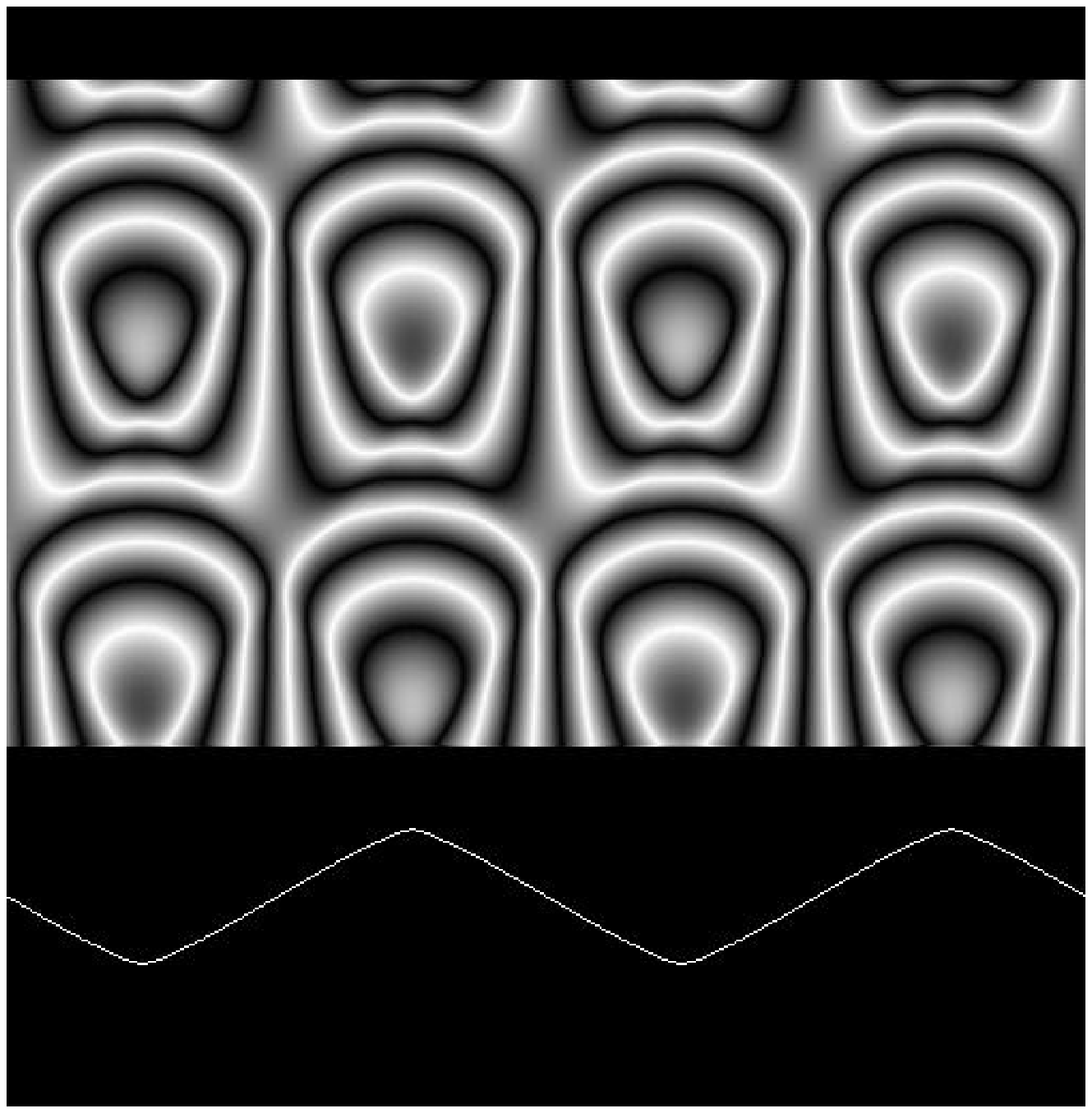}}&\resizebox{40mm}{!}{\includegraphics{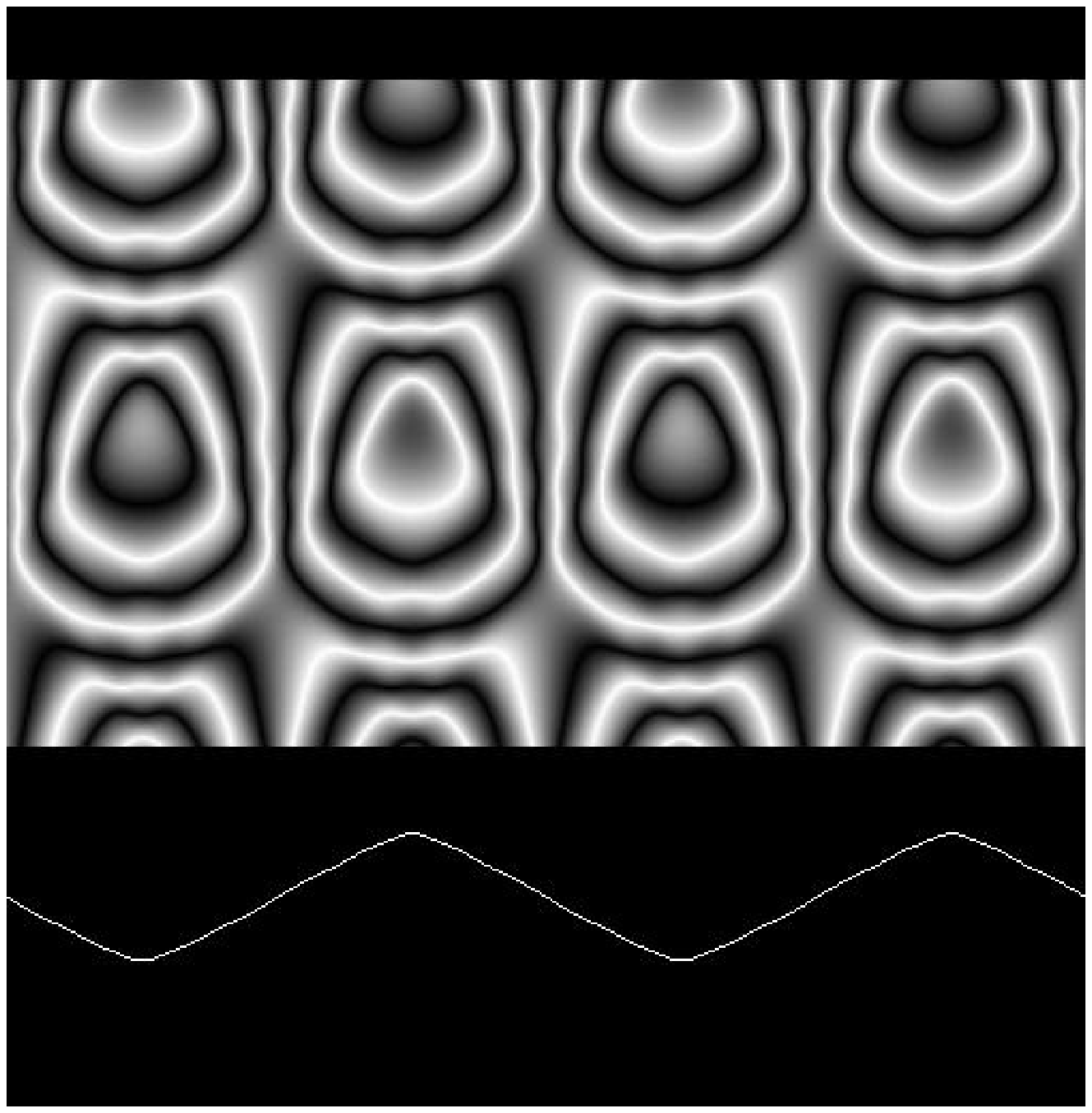}}\\
                \resizebox{40mm}{!}{\includegraphics{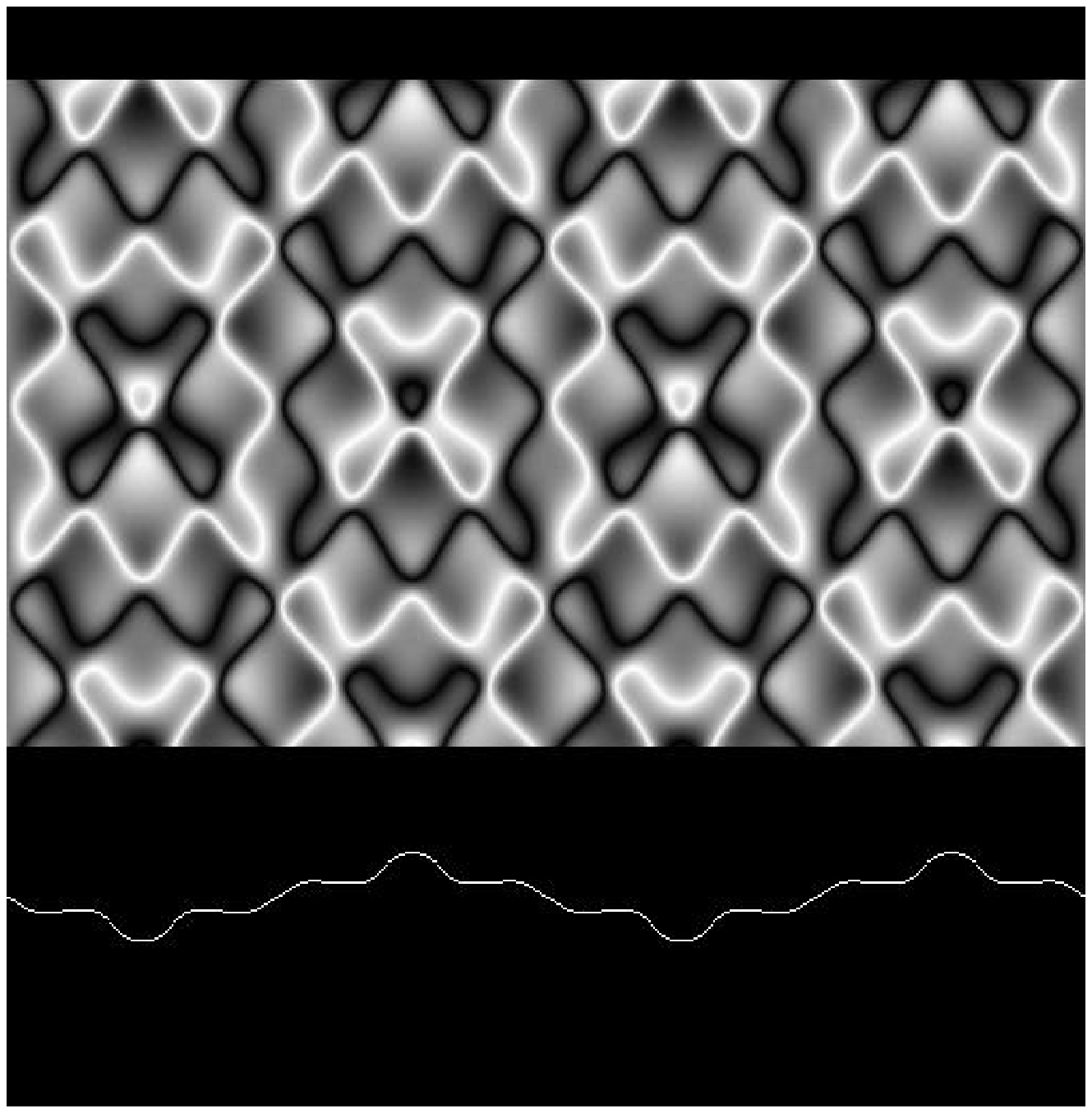}}&\resizebox{40mm}{!}{\includegraphics{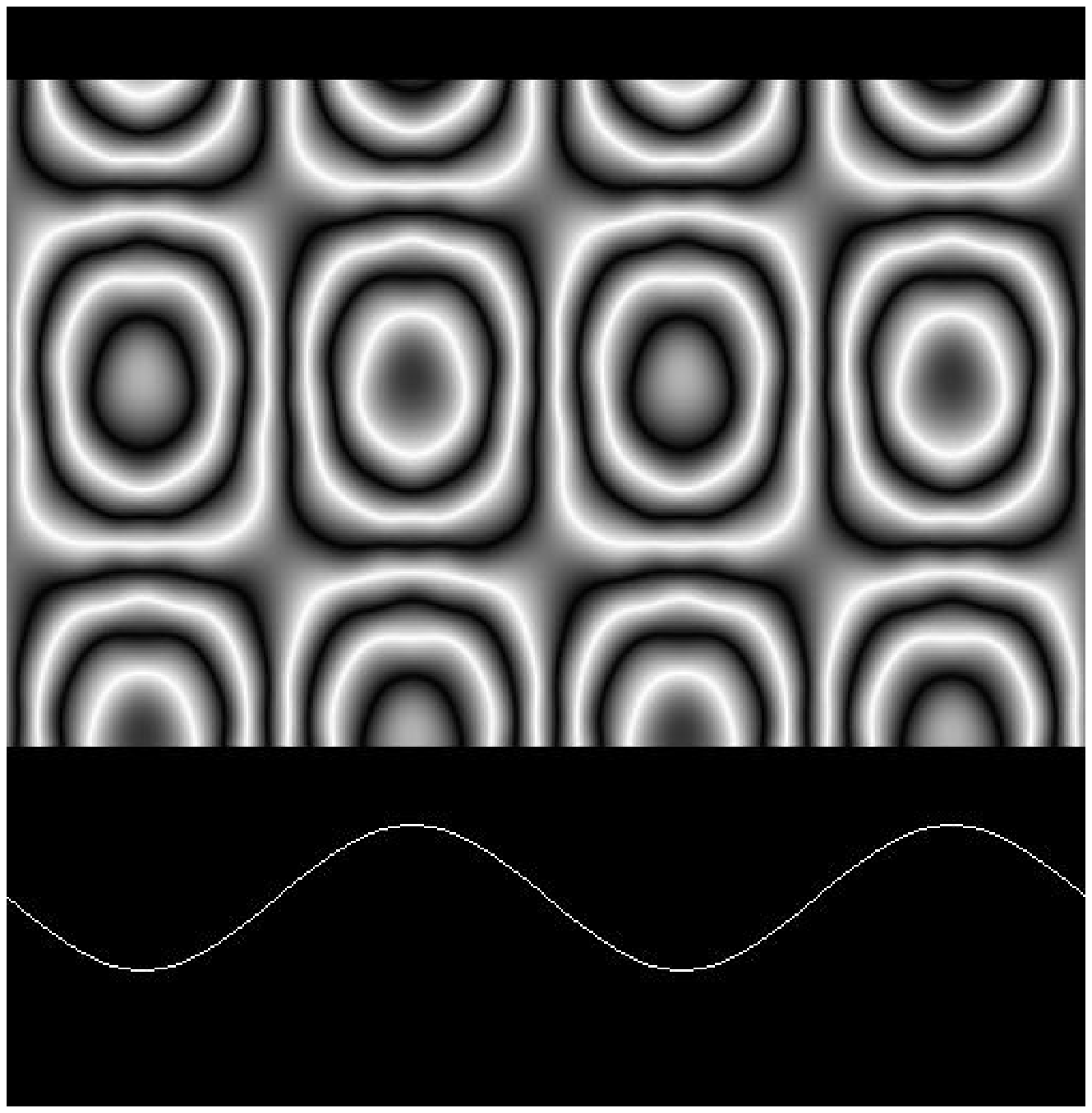}}
             \end{tabular}
         \caption{Time evolution of the shear wave structure for $\alpha=0.1$ for different time(time increasing column wise)
        with initial and final state almost identical. X-axis represents space and y-axis represents time slots}
         \label{fg1}
         \end{figure}

\begin{figure}
             \centering
             \begin{tabular}{cc}
                \resizebox{60mm}{!}{\includegraphics{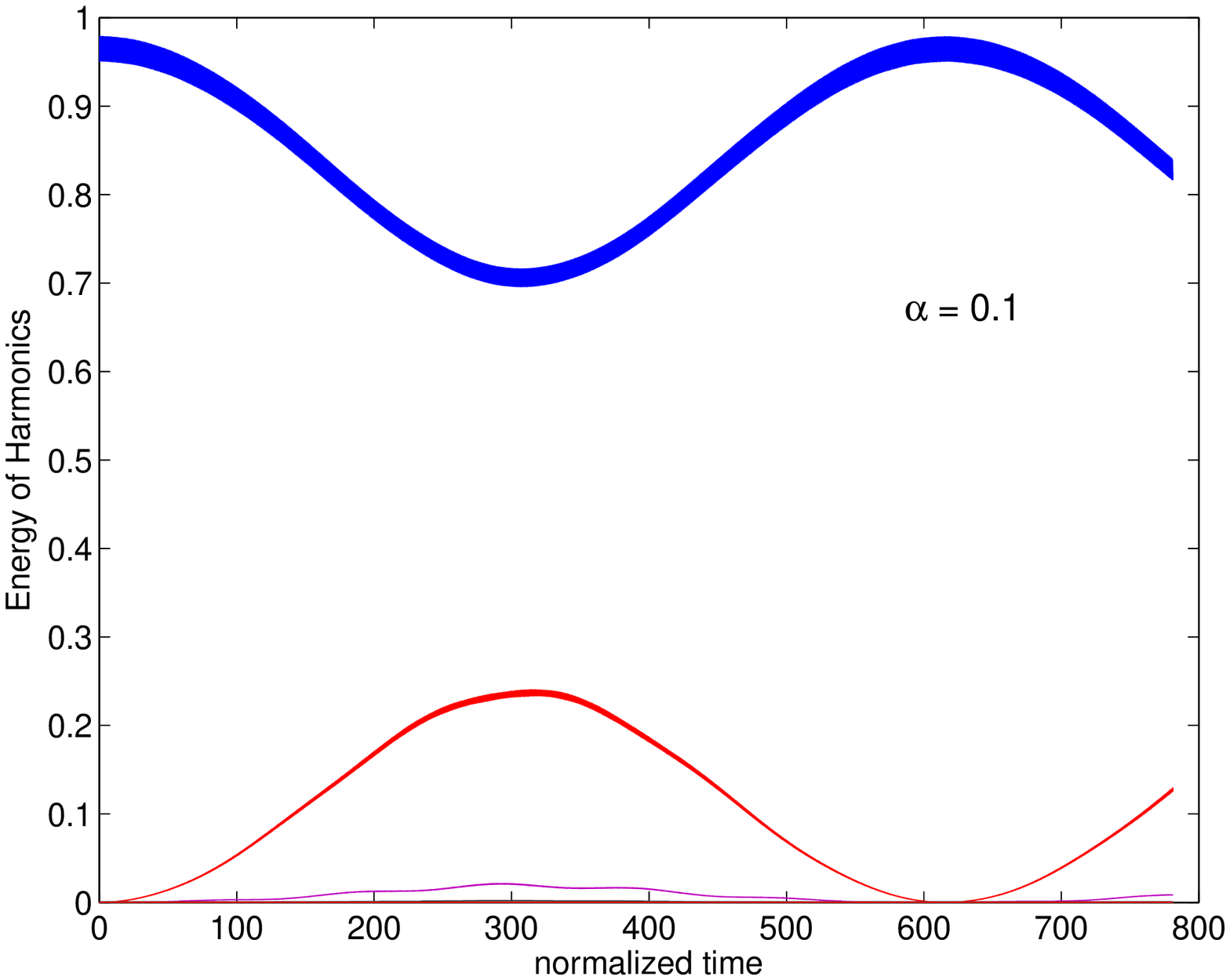}}&\resizebox{60mm}{!}{\includegraphics{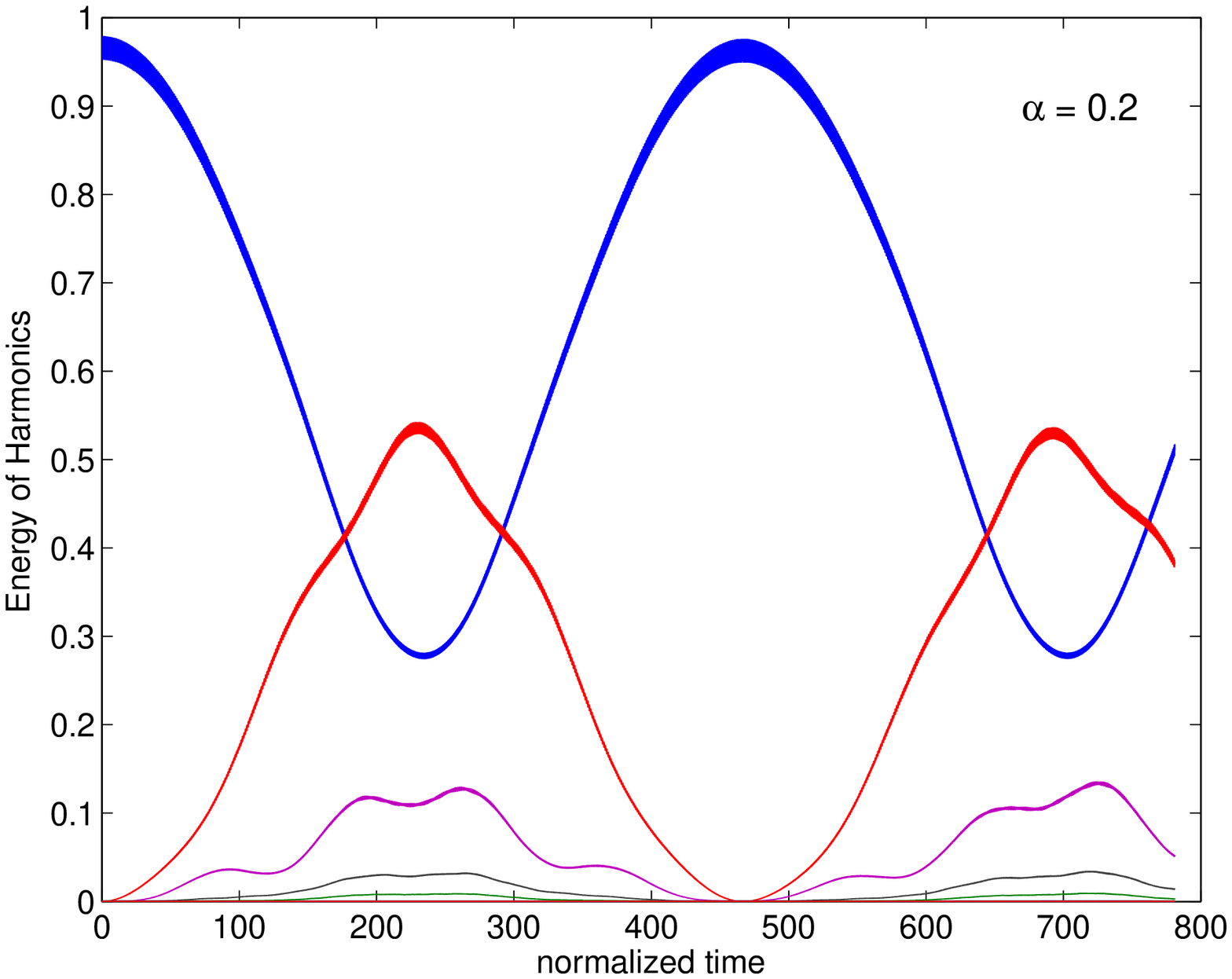}}\\
                \resizebox{60mm}{!}{\includegraphics{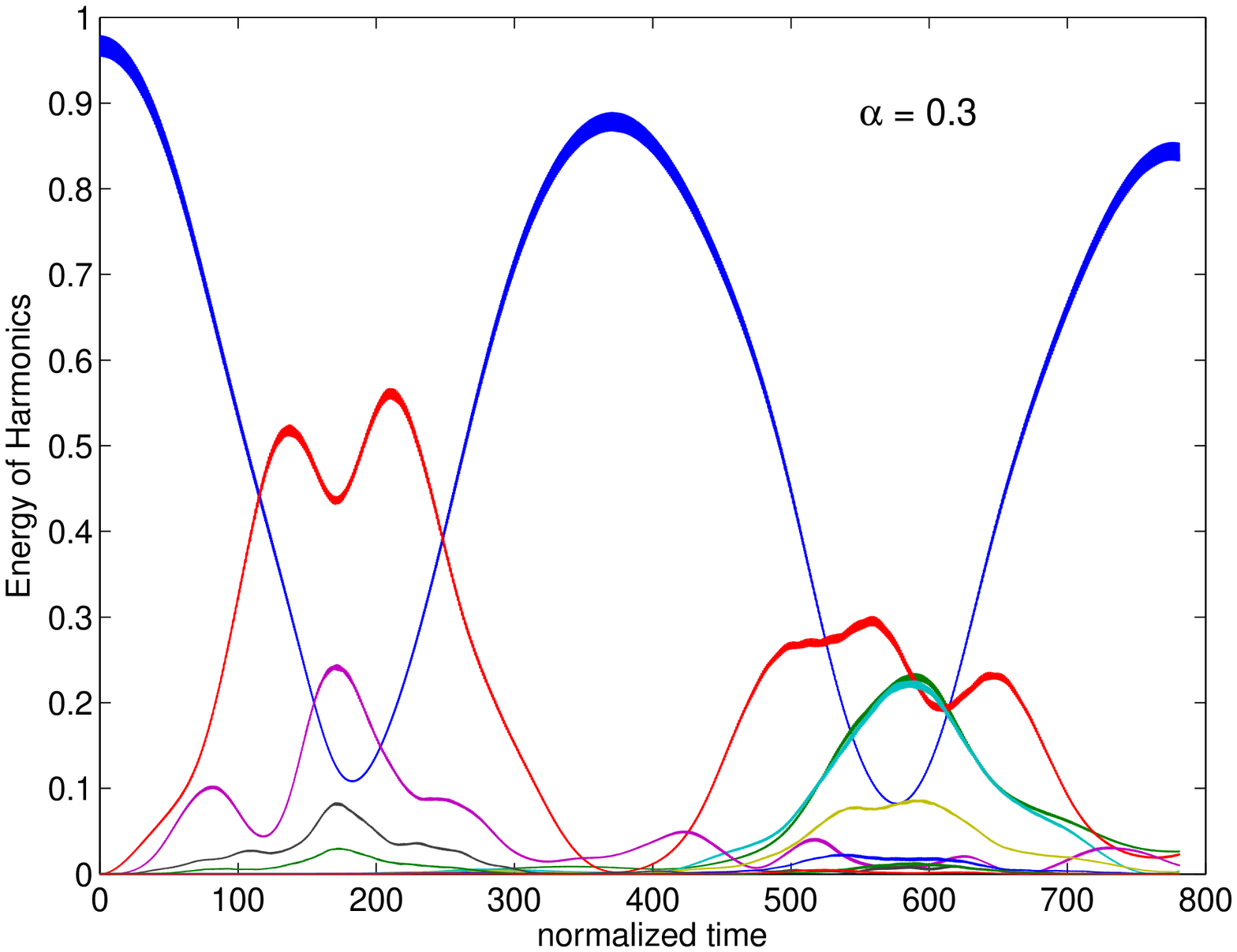}}&\resizebox{60mm}{!}{\includegraphics{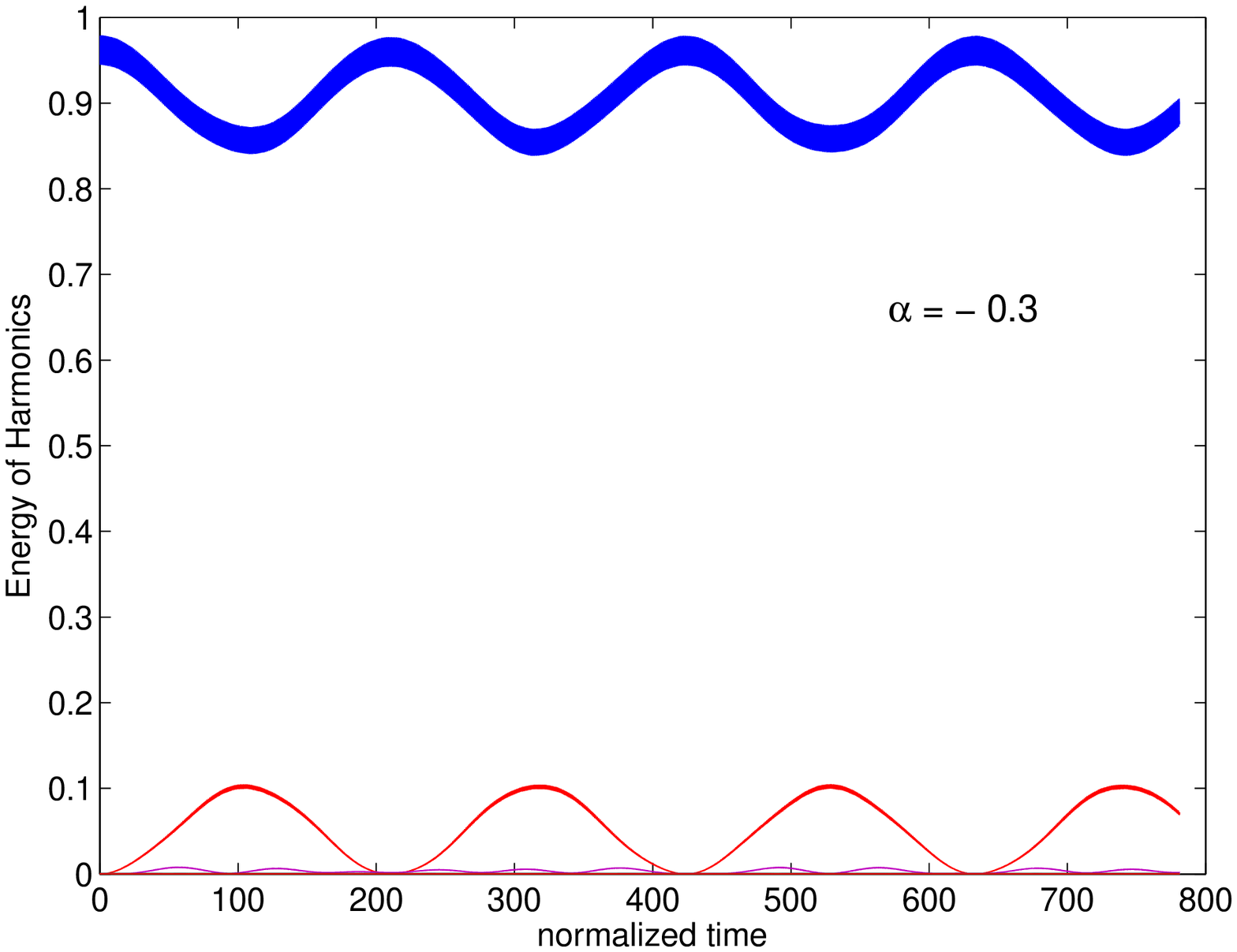}}\\
             \end{tabular}
             \caption{Energy of different harmonics is plotted against normalized time for different values of nonlinearity $\alpha$.
                       time is normalized by $c_{sh}/L$, where $L$ is the system length.
                       Blue line shows energy of fundamental harmonic and other red, brown, yellow etc show for different higher harmonics.
                       First three graphs show energy plots for three different positive values of $\alpha$. Hence, energy sharing to higher
                       harmonics is greater for $\alpha = 0.3$ than that of $0.1$ and $0.2$. Also recurrence is faster for larger values of
                       $\alpha$. Fourth graph is for shear thinning medium.}
            \label{fg2}
           \end{figure}

            Analytic solution is not straightforward therefore first we have studied this equation (\ref{nonlin}) numerically.
            Space is discretized with a minimum length $\Delta x$ and the equation is given by
         \bee
          \npar{2}{v}{t} =  \frac{1}{(\Delta x)^2} \left( v_{j+1} - 2 v_{j} + v_{j-1}\right) + \frac{\alpha}{(\Delta x)^4}
          \left[\left( v_{j+1} - v_j\right)^3 + \left( v_{j-1} - v_j \right)^3\right]
          \label{fd}
          \ene
          Next, we use mean value finite discretisation in time  and solved this equation using a standard software \cite{rudy,daux}.
          In this numerical study, minimum space and time step
          introduce two parameters namely $\beta = (\Delta t/\Delta x)^2$ and $\gamma = \alpha \beta/(\Delta x)^2$. In a typical
          example we have used in the numerical investigation $\Delta t = 0.03$, $\Delta x = 0.04$,
          $\gamma \approx 351.5 \alpha $ with $\alpha = 0.1$.  In order to ensure the numerical stability,
          one must satisfy the Courant-Friedrichs-Lewy  (CFL) condition. Initially (at t=0),
          we perturb the system with a pure sine wave and keep observing its changes in time. After few time steps, wave is seen to
          change into a triangular shape. As time goes on, the amplitude of the wave form diminishes and the periodicity of
          the  wave changes with time. Nonlinearity excites higher harmonics in the system and initial
          energy of fundamental mode is distributed through different harmonics(\ref{fg2}).
          The interesting feature is that after a large number of  time steps, all the
          higher modes starts to  disappear and initial energy of the system accumulates in
          the fundamental mode(initial sinusoidal perturbation as shown in Fig (\ref{fg1}).
          Therefore nonlinearity redistributes energy of the wave in different harmonics and they interact
          themselves and finally come back to its initial state. This feature of the solution
          reminds us the famous Fermi-Pasta -Ulam (FPU) problem in a completely different physical situation.
          Numerical analysis has also been carried out for shear thinning medium i.e when $\alpha$ is negative
          e.g, $\alpha = -0.3$. But, for this medium, energy sharing is negligible compared to that for shear thickening media and thus
          this case is not so interesting.

          It is shown that the solutions of eq.(\ref{nonlin}) retrace the initial condition
          and maintain its periodicity inspite of the strong nonlinearity present in the equation.
          We search a possible explanation transforming the discrete equation to the
          continuum limit keeping higher space derivative in the Taylor series expansion of the velocity function.
          With an appropriate scaling $t\rightarrow (\Delta x)t$ and $\psi
          \rightarrow (\Delta x) \psi\sqrt{1/3 \alpha}$  in equation(\ref{fd}) and
           defining a continuum field variable $\psi(x,t)$ \cite{dris}, as
          \bee
          \psi \equiv - \frac{v_t}{2(\Delta x)} + \frac{1}{2}\int_{0}^{v_x}\left(1 \pm (\Delta x)^2 \zeta^2 \right)^{1/2} d \zeta.
          \label{drs}
          \ene
          we have the equation
          \bee
           \fpar{\psi}{\tau} \pm 12 \psi^2 \fpar{\psi}{\xi} +\npar{3}{\psi}{\xi}=0.
           \label{mkdv}
          \ene
          Note here that to write the above mKdV equation we have kept only lowest order terms in $\Delta x$ after Taylor expansion,
          with $ \xi \equiv x - (\Delta x) t $ and $ \tau \equiv (\Delta x)^3 t /24$.
          The mkdV equation given above has well known periodic solutions of the form  $\sim cn^2$.
          This may be the case why we observe recurring solution in the  numerical investigation of  nonlinear shear wave equation.

           \section{Summary}\label{sec:sum} %
   In conclusion, we emphasize that the analysis of a nonlinear shear wave  has been
   done in a very simple macroscopic  situation in which nonlinearity, viscosity and
   velocity shear are treated on equal footing, resulting in an interesting solution of the governing equations.
   The nonlinear shear wave equation is numerically solved  by identifying it with the
   celebrated FPU problem with cubic nonlinearity. We could show that for the shear thickening situation nonlinear shear waves redistribute their energy
   and finally come back to the initial state while for a shear thinning medium energy distribution is negligibly small.
   Therefore the propagation of transverse shear waves  is an important technique
   for characterization of strongly coupled Yukawa fluids.
    Our solution is well applicable to other branches of strongly coupled media such as
    neutral viscoelastic fluids ranging from polymer solutions, biological fluids like blood,
    pulmonary liquids, human soft tissues, to  magma fluids.  In such cases linear response is controlled by imposing small
    amplitude displacements.  However, it is
   interesting and also desirable to extend such investigations  to models with nonlinear constitutive equations leading to nonlinear
   physics.  Such methods have the potential to
   characterize  nonlinear viscoelastic properties  at low stress levels that
   are typical in biological conditions.  The important contribution of this work is to form
   and solve the evolution equation for nonlinear shear waves.
   This type of solution represents a new class of nonlinear solutions that may arise in a manifold of
   similar physical situations.
   The dependence of the time period of recurrence of periodic solutions
   on the nonlinearity parameter can enable the characterization of non-Newtonian properties possible.
   We believe that more investigations in this direction will enrich our knowledge on the physical understanding
   of strongly coupled media.
%

%

\begin{thebibliography}{10}%
\makeatletter
\providecommand \@ifxundefined [1]{%
 \ifx #1\undefined \expandafter \@firstoftwo
 \else \expandafter \@secondoftwo
\fi
}%
\providecommand \@ifnum [1]{%
 \ifnum #1\expandafter \@firstoftwo
 \else \expandafter \@secondoftwo
\fi
}%
\providecommand \enquote [1]{``#1''}%
\providecommand \bibnamefont  [1]{#1}%
\providecommand \bibfnamefont [1]{#1}%
\providecommand \citenamefont [1]{#1}%
\providecommand\href[0]{\@sanitize\@href}%
\providecommand\@href[1]{\endgroup\@@startlink{#1}\endgroup\@@href}%
\providecommand\@@href[1]{#1\@@endlink}%
\providecommand \@sanitize [0]{\begingroup\catcode`\&12\catcode`\#12\relax}%
\@ifxundefined \pdfoutput {\@firstoftwo}{%
 \@ifnum{\z@=\pdfoutput}{\@firstoftwo}{\@secondoftwo}%
}{%
 \providecommand\@@startlink[1]{\leavevmode}%
 \providecommand\@@endlink[0]{}%
}{%
 \providecommand\@@startlink[1]{%
  \leavevmode
  \pdfstartlink
   attr{/Border[0 0 1 ]/H/I/C[0 1 1]}%
   user{/Subtype/Link/A<</Type/Action/S/URI/URI(#1)>>}%
  \relax
 }%
 \providecommand\@@endlink[0]{\pdfendlink}%
}%
\providecommand \url  [0]{\begingroup\@sanitize \@url }%
\providecommand \@url [1]{\endgroup\@href {#1}{\urlprefix}}%
\providecommand \urlprefix [0]{URL }%
\providecommand \Eprint[0]{\href }%
\@ifxundefined \urlstyle {%
  \providecommand \doi [1]{doi:\discretionary{}{}{}#1}%
}{%
  \providecommand \doi [0]{doi:\discretionary{}{}{}\begingroup
  \urlstyle{rm}\Url }%
}%
\providecommand \doibase [0]{http://dx.doi.org/}%
\providecommand \Doi[1]{\href{\doibase#1}}%
\providecommand \bibAnnote [3]{%
  \BibitemShut{#1}%
  \begin{quotation}\noindent
    \textsc{Key:}\ #2\\\textsc{Annotation:}\ #3%
  \end{quotation}%
}%
\providecommand \bibAnnoteFile [2]{%
  \IfFileExists{#2}{\bibAnnote {#1} {#2} {\input{#2}}}{}%
}%
\providecommand \typeout [0]{\immediate \write \m@ne }%
\providecommand \selectlanguage [0]{\@gobble}%
\providecommand \bibinfo [0]{\@secondoftwo}%
\providecommand \bibfield [0]{\@secondoftwo}%
\providecommand \translation [1]{[#1]}%
\providecommand \BibitemOpen[0]{}%
\providecommand \bibitemStop [0]{}%
\providecommand \bibitemNoStop [0]{.\EOS\space}%
\providecommand \EOS [0]{\spacefactor3000\relax}%
\providecommand \BibitemShut [1]{\csname bibitem#1\endcsname}%
\bibitem{shma}%
  \BibitemOpen
  \bibfield{author}{%
  \bibinfo {author} {\bibfnamefont{P.~K.}\ \bibnamefont{Shukla}}\ and\ \bibinfo
  {author} {\bibfnamefont{A.~A.}\ \bibnamefont{Mamun}},\ }%
  \emph{\bibinfo {title} {Introduction to Dusty Plasma Physics}}\ (\bibinfo
  {publisher} {Institute of Physics Publication},\ \bibinfo {address}
  {Bristol},\ \bibinfo {year} {2002})%
  \bibAnnoteFile{NoStop}{shma}%
\bibitem{ichi}%
  \BibitemOpen
  \bibfield{author}{%
  \bibinfo {author} {\bibfnamefont{S.}~\bibnamefont{Ichimaru}},\ }%
  \bibfield{journal}{%
  \bibinfo {journal} {Rev. Mod. Phys.}\ }%
  \textbf{\bibinfo {volume} {54}},\ \bibinfo {pages} {1017} (\bibinfo {year}
  {1982})%
  \bibAnnoteFile{NoStop}{ichi}%
\bibitem{pkmt}%
  \BibitemOpen
  \bibfield{author}{%
  \bibinfo {author} {\bibfnamefont{P.~K.}\ \bibnamefont{Shukla}}, \bibinfo
  {author} {\bibfnamefont{M.}~\bibnamefont{Torney}}, \bibinfo {author}
  {\bibfnamefont{R.}~\bibnamefont{Bingham}},\ and\ \bibinfo {author}
  {\bibfnamefont{G.~E.}\ \bibnamefont{Morfill}},\ }%
  \bibfield{journal}{%
  \bibinfo {journal} {Phys. Scripta.}\ }%
  \textbf{\bibinfo {volume} {67}},\ \bibinfo {pages} {350} (\bibinfo {year}
  {2003})%
  \bibAnnoteFile{NoStop}{pkmt}%
\bibitem{msjn}%
  \BibitemOpen
  \bibfield{author}{%
  \bibinfo {author} {\bibfnamefont{M.~S.}\ \bibnamefont{Janaki}}, \bibinfo
  {author} {\bibfnamefont{N.}~\bibnamefont{Chakrabarti}},\ and\ \bibinfo
  {author} {\bibfnamefont{D.}~\bibnamefont{Banerjee}},\ }%
  \bibfield{journal}{%
  \bibinfo {journal} {Phys. Plasmas}\ }%
  \textbf{\bibinfo {volume} {18}},\ \bibinfo {pages} {012901} (\bibinfo {year}
  {2011})%
  \bibAnnoteFile{NoStop}{msjn}%
\bibitem{pkaw}%
  \BibitemOpen
  \bibfield{author}{%
  \bibinfo {author} {\bibfnamefont{P.}~\bibnamefont{Kaw}}\ and\ \bibinfo
  {author} {\bibfnamefont{A.}~\bibnamefont{Sen}},\ }%
  \bibfield{journal}{%
  \bibinfo {journal} {Phys. Plasmas}\ }%
  \textbf{\bibinfo {volume} {5}},\ \bibinfo {pages} {3552} (\bibinfo {year}
  {1998})%
  \bibAnnoteFile{NoStop}{pkaw}%
\bibitem{mash}%
  \BibitemOpen
  \bibfield{author}{%
  \bibinfo {author} {\bibfnamefont{A.~A.}\ \bibnamefont{Mamun}}\ and\ \bibinfo
  {author} {\bibfnamefont{P.~K.}\ \bibnamefont{Shukla}},\ }%
  \bibfield{journal}{%
  \bibinfo {journal} {Euro. Phys. Letter}\ }%
  \textbf{\bibinfo {volume} {87}},\ \bibinfo {pages} {55001} (\bibinfo {year}
  {2009})%
  \bibAnnoteFile{NoStop}{mash}%
\bibitem{banj}%
  \BibitemOpen
  \bibfield{author}{%
  \bibinfo {author} {\bibfnamefont{D.}~\bibnamefont{Banerjee}}, \bibinfo
  {author} {\bibfnamefont{M.~S.}\ \bibnamefont{janaki}},\ and\ \bibinfo
  {author} {\bibfnamefont{N.}~\bibnamefont{Chakrabarti}},\ }%
  \bibfield{journal}{%
  \bibinfo {journal} {Phys. Plasmas}\ }%
  \textbf{\bibinfo {volume} {17}},\ \bibinfo {pages} {113708} (\bibinfo {year}
  {2010})%
  \bibAnnoteFile{NoStop}{banj}%
\bibitem{pram}%
  \BibitemOpen
  \bibfield{author}{%
  \bibinfo {author} {\bibfnamefont{J.}~\bibnamefont{Pramanik}}, \bibinfo
  {author} {\bibfnamefont{G.}~\bibnamefont{Prasad}}, \bibinfo {author}
  {\bibfnamefont{A.}~\bibnamefont{Sen}},\ and\ \bibinfo {author}
  {\bibfnamefont{P.}~\bibnamefont{Kaw}},\ }%
  \bibfield{journal}{%
  \bibinfo {journal} {Phys. Rev. Lett.}\ }%
  \textbf{\bibinfo {volume} {88}},\ \bibinfo {pages} {17500} (\bibinfo {year}
  {2002})%
  \bibAnnoteFile{NoStop}{pram}%
\bibitem{nose}%
  \BibitemOpen
  \bibfield{author}{%
  \bibinfo {author} {\bibfnamefont{V.}~\bibnamefont{Nosenko}}\ and\ \bibinfo
  {author} {\bibfnamefont{J.}~\bibnamefont{Goree}},\ }%
  \bibfield{journal}{%
  \bibinfo {journal} {Phys.\ Rev.\ Lett.}\ }%
  \textbf{\bibinfo {volume} {93}},\ \bibinfo {pages} {155004} (\bibinfo {year}
  {2004})%
  \bibAnnoteFile{NoStop}{nose}%
\bibitem{ivle}%
  \BibitemOpen
  \bibfield{author}{%
  \bibinfo {author} {\bibfnamefont{A.~V.}\ \bibnamefont{Ivlev}}, \bibinfo
  {author} {\bibfnamefont{V.}~\bibnamefont{Steinberg}}, \bibinfo {author}
  {\bibfnamefont{R.}~\bibnamefont{Kompaneets}}, \bibinfo {author}
  {\bibfnamefont{H.}~\bibnamefont{H$\ddot{o}$fner}}, \bibinfo {author}
  {\bibfnamefont{I.}~\bibnamefont{Sidorenko}},\ and\ \bibinfo {author}
  {\bibfnamefont{G.~E.}\ \bibnamefont{Morfill}},\ }%
  \bibfield{journal}{%
  \bibinfo {journal} {Phys. Rev. Lett.}\ }%
  \textbf{\bibinfo {volume} {98}},\ \bibinfo {pages} {145003} (\bibinfo {year}
  {2007})%
  \bibAnnoteFile{NoStop}{ivle}%
\bibitem{jpps}%
  \BibitemOpen
  \bibfield{author}{%
  \bibinfo {author} {\bibfnamefont{A.~V.}\ \bibnamefont{Gavrikov}}, \bibinfo
  {author} {\bibfnamefont{D.~N.}\ \bibnamefont{Goranskaya}}, \bibinfo {author}
  {\bibfnamefont{A.~S.}\ \bibnamefont{Ivanov}}, \bibinfo {author}
  {\bibfnamefont{O.~F.}\ \bibnamefont{Petrov}}, \bibinfo {author}
  {\bibfnamefont{R.~A.}\ \bibnamefont{Timirkhanov}}, \bibinfo {author}
  {\bibfnamefont{N.~A.}\ \bibnamefont{Vorona}},\ and\ \bibinfo {author}
  {\bibfnamefont{V.~E.}\ \bibnamefont{Fortov}},\ }%
  \bibfield{journal}{%
  \bibinfo {journal} {J. Plasma Physics}\ }%
  \textbf{\bibinfo {volume} {76}},\ \bibinfo {pages} {579} (\bibinfo {year}
  {2010})%
  \bibAnnoteFile{NoStop}{jpps}%
\bibitem{sora}%
  \BibitemOpen
  \bibfield{author}{%
  \bibinfo {author} {\bibfnamefont{G.}~\bibnamefont{Sorasio}}, \bibinfo
  {author} {\bibfnamefont{P.~K.}\ \bibnamefont{Shukla}},\ and\ \bibinfo
  {author} {\bibfnamefont{D.~P.}\ \bibnamefont{Resendes}},\ }%
  \bibfield{journal}{%
  \bibinfo {journal} {New. J. Phys.}\ }%
  \textbf{\bibinfo {volume} {5}},\ \bibinfo {pages} {81} (\bibinfo {year}
  {2003})%
  \bibAnnoteFile{NoStop}{sora}%
\bibitem{mamu}%
  \BibitemOpen
  \bibfield{author}{%
  \bibinfo {author} {\bibfnamefont{A.~A.}\ \bibnamefont{Mamun}}, \bibinfo
  {author} {\bibfnamefont{P.~K.}\ \bibnamefont{Shukla}},\ and\ \bibinfo
  {author} {\bibfnamefont{T.}~\bibnamefont{Farid}},\ }%
  \bibfield{journal}{%
  \bibinfo {journal} {Phys. Plasmas.}\ }%
  \textbf{\bibinfo {volume} {7}},\ \bibinfo {pages} {2329} (\bibinfo {year}
  {2000})%
  \bibAnnoteFile{NoStop}{mamu}%
\bibitem{bird}%
  \BibitemOpen
  \bibfield{author}{%
  \bibinfo {author} {\bibfnamefont{R.~B.}\ \bibnamefont{Bird}}, \bibinfo
  {author} {\bibfnamefont{R.~C.}\ \bibnamefont{Armstrong}},\ and\ \bibinfo
  {author} {\bibfnamefont{O.}~\bibnamefont{Hassger}},\ }%
  \emph{\bibinfo {title} {Dynamics of polymeric liquids Vol. 1}}\ (\bibinfo
  {publisher} {Wiley},\ \bibinfo {address} {New York},\ \bibinfo {year}
  {1977})%
  \bibAnnoteFile{NoStop}{bird}%
\bibitem{bisw}%
  \BibitemOpen
  \bibfield{author}{%
  \bibinfo {author} {\bibfnamefont{E.~E.~B.}\ \bibnamefont{White}}, \bibinfo
  {author} {\bibfnamefont{M.}~\bibnamefont{Chellamuthu}},\ and\ \bibinfo
  {author} {\bibfnamefont{J.~P.}\ \bibnamefont{Rothstein}},\ }%
  \bibfield{journal}{%
  \bibinfo {journal} {Rheol Acta}\ }%
  \textbf{\bibinfo {volume} {49}},\ \bibinfo {pages} {119} (\bibinfo {year}
  {2010})%
  \bibAnnoteFile{NoStop}{bisw}%
\bibitem{thur}%
  \BibitemOpen
  \bibfield{author}{%
  \bibinfo {author} {\bibfnamefont{G.}~\bibnamefont{Thurston}},\ }%
  \bibfield{journal}{%
  \bibinfo {journal} {Biophys. J.}\ }%
  \textbf{\bibinfo {volume} {12}},\ \bibinfo {pages} {1205} (\bibinfo {year}
  {1972})%
  \bibAnnoteFile{NoStop}{thur}%
\bibitem{fpu}%
  \BibitemOpen
  \bibfield{author}{%
  \bibinfo {author} {\bibfnamefont{E.}~\bibnamefont{Fermi}}, \bibinfo {author}
  {\bibfnamefont{J.}~\bibnamefont{Pasta}},\ and\ \bibinfo {author}
  {\bibfnamefont{S.}~\bibnamefont{Ulam}},\ }%
  \emph{\bibinfo {title} {Studies of Nonlinear Problem I}}\ (\bibinfo
  {publisher} {Los Alomas Report},\ \bibinfo {address} {LA-1940},\ \bibinfo
  {year} {1955})%
  \bibAnnoteFile{NoStop}{fpu}%
\bibitem{fren}%
  \BibitemOpen
  \bibfield{author}{%
  \bibinfo {author} {\bibfnamefont{Y.}~\bibnamefont{Frenkel}},\ }%
  \emph{\bibinfo {title} {Kinetic Theory of Liquids}}\ (\bibinfo {publisher}
  {Clarendon},\ \bibinfo {address} {Oxford},\ \bibinfo {year} {1946})%
  \bibAnnoteFile{NoStop}{fren}%
\bibitem{rudy}%
  \BibitemOpen
  \bibfield{author}{%
  \bibinfo {author} {\bibfnamefont{R.}~\bibnamefont{Rucker{\it{ et. al.}}}},\
  }%
  \bibfield{journal}{%
  \bibinfo {journal} {http://www.rudyrucker.com/capow}\ }%
  \textbf{\bibinfo {volume} {software}},\ \bibinfo {pages} {downloaded}
  (\bibinfo {year} {2007})%
  \bibAnnoteFile{NoStop}{rudy}%
\bibitem{daux}%
  \BibitemOpen
  \bibfield{author}{%
  \bibinfo {author} {\bibfnamefont{T.}~\bibnamefont{Dauxois}}, \bibinfo
  {author} {\bibfnamefont{M.}~\bibnamefont{Peyrard}},\ and\ \bibinfo {author}
  {\bibfnamefont{S.}~\bibnamefont{Ruffo}},\ }%
  \bibfield{journal}{%
  \bibinfo {journal} {European. J. Phys.}\ }%
  \textbf{\bibinfo {volume} {26}},\ \bibinfo {pages} {S3} (\bibinfo {year}
  {2005})%
  \bibAnnoteFile{NoStop}{daux}%
\bibitem{dris}%
  \BibitemOpen
  \bibfield{author}{%
  \bibinfo {author} {\bibfnamefont{C.~F.}\ \bibnamefont{Driscoll}}\ and\
  \bibinfo {author} {\bibfnamefont{T.~M.}\ \bibnamefont{O'Neil}},\ }%
  \bibfield{journal}{%
  \bibinfo {journal} {Phys. Rev. Lett.}\ }%
  \textbf{\bibinfo {volume} {37}},\ \bibinfo {pages} {89} (\bibinfo {year}
  {1976})%
  \bibAnnoteFile{NoStop}{dris}%
\end{thebibliography}
   \end{document}